\documentclass[
aps,
prd,
twocolumn,
10pt,
preprintnumbers,
notitlepage,
amssymb,
groupedaddress,
nofootinbib,
noeprint
]{revtex4-2}
\pdfoutput=1 
\usepackage[utf8]{inputenc}
\usepackage[english]{babel}
\usepackage{float}
\usepackage{color}
\usepackage{booktabs}
\usepackage{amsmath,amsfonts,amssymb,mathrsfs,amsthm}
\usepackage{nicefrac}
\usepackage{mathtools}
\usepackage{url}
\usepackage[hyperindex,breaklinks]{hyperref}
\begin{document}
\title{Naked singularities in quadratic $f(R)$ gravity}
\preprint{FR-PHENO-2020-004}
\author{Enrique Hernand\'ez-Lorenzo}
\author{Christian F.~Steinwachs}
\email{christian.steinwachs@physik.uni-freiburg.de}
\affiliation{Physikalisches Institut, Albert-Ludwigs-Universit\"at Freiburg,\\
Hermann-Herder-Stra\ss e~3, 79104 Freiburg, Germany}
%
%%********************************************%
%
\begin{abstract}
We find new static, spherically symmetric, and asymptotically flat vacuum solutions without horizon in Starobinsky's quadratic $f(R)$ gravity.
We systematically classify these solutions by an asymptotic analysis around the origin and find seven different integer Frobenius families.
We numerically solve the exact equations of motion by a double-shooting method and specify boundary conditions by matching the numerical solution to the analytic solution of the linearised field equations in the weak field regime.
We find that all integer Frobenius families can be connected to asymptotically flat solutions and trace out lines in the parameter space, allowing to ultimately relate all free parameters to the total mass at infinity.   
\end{abstract}
%
%%%%%%%%%%%%%%%%%%%%%%%%%%%%%%%%%%%%%%%					  
%\pacs{}		  
\maketitle								  
%%%%%%%%%%%%%%%%%%%%%%%%%%%%%%%%%%%%%%%
%
%%%%%%%%%%%%%%%%%%%%%%%%%%%%%%%%%%%%%%%
\section{Introduction}
%%%%%%%%%%%%%%%%%%%%%%%%%%%%%%%%%%%%%%%
%
Theories incorporating geometric modifications of General Relativity (GR) are highly relevant in the context of quantum gravity and cosmology.
In particular, $f(R)$ modifications of GR give rise to phenomenologically vital cosmological models for the early and late time acceleration of the Universe \cite{Sotiriou2010,DeFelice2010,Clifton2012}, of which Starobinsky's quadratic  $f(R)=R+R^2$ model of inflation \cite{Starobinsky1980} is the most relevant one and favoured by CMB data \cite{Akrami2018}.
The higher derivative structure of $f(R)$ gravity leads to an additional scalar degree of freedom, the \textit{scalaron}, made manifest in the classically equivalent scalar-tensor representation.
In \cite{Ruf2018c,Ohta2018a}, it was shown that this classical equivalence also extends to the quantum level in a similar way as the on-shell quantum equivalence between different field parametrizations in scalar-tensor theories found in \cite{Kamenshchik2015}.
In contrast to higher-derivative theories such as Quadratic Gravity (QG), the degeneracy structure of $f(R)$ gravity ensures the absence of the Ostrogradski instability \cite{Ostrogradsky1850, Stelle1978, Woodard2007,Ruf2018}. 

In order to investigate the construction of modified theories of gravity, it is important to compare the structure of solutions to that of GR. In general, as for GR, analytical solutions are only available for a high degree of symmetry.
However, in contrast to GR, in general no uniqueness theorems are available in higher derivative theories of gravity, see e.g.~\cite{Schmidt2012} and references therein.
Nevertheless, in quadratic gravity (QG) there is a ``trace no-hair theorem'' \cite{Nelson2010}, which implies that static asymptotically flat vacuum solutions with a horizon must have vanishing Ricci scalar $R$.
Therefore, in the search for black hole solutions, the $R^2$ term does not contribute to the equations of motion and the black hole analysis in QG effectively reduces to that in Einstein-Weyl gravity (EWG) \cite{Lu2015,Lue2015,Perkins2016-09,Pravda2017,Podolsky2018,Bonanno2019,Podolsky2020}.
In contrast, we do not search for black hole solutions, but instead focus on solutions without a horizon.
Solutions without a horizon but a singular geometry feature naked singularities and might have interesting implications for the quantum information paradox \cite{Holdom2017} and might serve as candidates for dark matter \cite{Aydemir2020}.

In this letter, we systematically investigate and classify static, spherically symmetric and asymptotically flat vacuum solutions in Starobinsky's quadratic $f(R)$ theory by a combined analytic and numerical approach.
We first analytically perform the asymptotic analyses in the near origin regime and in the linearised weak field regime and then connect the two asymptotic regimes by numerically solving the full non-linear field equations by a ``double-shooting'' algorithm.
In particular, we do not make any additional simplifying assumptions about the curvature or the metric functions in the spherically symmetric ansatz of the line element, which allows us to screen the full space of solutions.\vspace{2mm}
%  
%%%%%%%%%%%%%%%%%%%%%%%%%%%%%%%%%%%%%%%
\section{Starobinsky's quadratic $f(R)$ model}
\label{Sec:fRGravity}
%%%%%%%%%%%%%%%%%%%%%%%%%%%%%%%%%%%%%%%
%
The $f(R)$ action in four spacetime dimensions reads
\begin{align}
S[g]=\int\mathop{}\!\mathrm{d}^{4}x\, \sqrt{-g}\,f(R).\label{fRAct}
\end{align}
Derivatives of $f$ with respect to $R$ are denoted by ${f_{n}:=\partial^{n}f(R)/\partial R^n}$ and the equations of motion (EOM) of $f(R)$ gravity read $\mathscr{E}^{\mu\nu}=0$ with the extremal tensor
\begin{align}
\mathscr{E}^{\mu\nu}:={}& 
\left(g^{\mu\nu}\Box-\nabla^{\mu}\nabla^{\nu}\right) f_1+R^{\mu\nu}f_1-\frac{1}{2}g^{\mu\nu}f.\label{Extremal}
\end{align}
The trace of the extremal tensor is defined as
\begin{align}
\mathscr{E}:= g_{\mu\nu}\mathscr{E}^{\mu\nu}=3\Box f_1+Rf_1-2f.\label{ExtremalTrace}
\end{align}
The invariance of the action \eqref{fRAct} under diffeomorphisms implies the generalized Bianchi identity
\begin{align}
\nabla_{\mu}\mathscr{E}^{\mu\nu}=0.\label{Bianchi}
\end{align}
The fourth-order nature of the field equations implies that in addition to the massless spin-two graviton, $f(R)$ gravity propagates the massive spin-zero scalaron.
We consider the Starobinsky model defined by \eqref{fRAct} with 
\begin{align}
f(R)=\frac{M_{\mathrm{P}}^2}{2}\left(R+\frac{R^2}{6M_{\mathrm{S}}^2}\right).\label{Staract}
\end{align}
The two mass scales (in natural units $c=\hbar=1$) are set by the reduced Planck mass $M_{\mathrm{P}}=1/\sqrt{8\pi G_{\mathrm{N}}}$ and the scalaron mass $M_{\mathrm{S}}^2\geq0$.
The inequality ensures that the scalaron is not a tachyon and $M_{\mathrm{S}}=0$ corresponds to the induced gravity limit $M_{\mathrm{P}}\to0$ of \eqref{Staract} resulting in the scale invariant $R^2$ model.
In case of a scalaron-driven inflationary phase of the early universe, $M_{\mathrm{S}}$ is constrained by Cosmic Microwave Background (CMB) data \cite{Akrami2018},
\begin{align}
M_{\mathrm{S}}\approx 10^{-5}M_{\mathrm{P}}\approx 10^{13}\mathrm{GeV}.\label{CMBBoundMS}
\end{align} 

%%%%%%%%%%%%%%%%%%%%%%%%%%%%%%%%%%%%%%%
\section{Static spherically symmetric metric ansatz}
\label{Sec:SSSMetric}
%%%%%%%%%%%%%%%%%%%%%%%%%%%%%%%%%%%%%%%

In Schwarzschild coordinates $(t,r,\vartheta,\phi)$, the static and spherically symmetric line element is parametrized as
\begin{align}
\mathrm{d}s^2=-B(r)\mathrm{d}t^2+A(r)\mathrm{d}r^2+r^2\mathrm{d}\Omega^2_{(2)}.\label{LESSS}
\end{align}
Here, ${\mathrm{d}\Omega^2_{(2)}=\mathrm{d}\vartheta^2+\sin^2\vartheta\mathrm{d}\phi^2}$ is the line element of the unit two-sphere and $A(r)$ and $B(r)$ are arbitrary functions of the radius $r$.
Only two of the non-zero components $\mathscr{E}^{tt}$, $\mathscr{E}^{rr}$, $\mathscr{E}^{\vartheta\vartheta}$ and $\mathscr{E}^{\phi\phi}$ are independent due to the redundancy  $\mathscr{E}^{\phi\phi}=\sin^2\vartheta\mathscr{E}^{\vartheta\vartheta}$ and the generalized Bianchi identity \eqref{Bianchi}.
We work with the EOMs $\mathscr{E}^{tt}=0$ and $\mathscr{E}^{rr}=0$, which have the functional dependencies
\begin{align}
\mathscr{E}^{tt}(B_4,B_3,B_2,B_1,B,A_3, A_2,A_1,A, r;M_{\mathrm{S}})={}&0,\label{FuncShape1}\\
\mathscr{E}^{rr}(B_3,B_2,B_1,B,A_2,A_1,A,r;M_{\mathrm{S}})={}&0.\label{FuncShape2}
\end{align}
The $n$th derivatives of $B$ and $A$ with respect to $r$ are denoted by $B_n$ and $A_n$. The explicit expressions for $\mathscr{E}^{tt}$ and $\mathscr{E}^{rr}$ are provided in \eqref{EttExpl} and \eqref{ErrExpl}.
In the form \eqref{FuncShape1}, \eqref{FuncShape2}, the differential order of this system is not transparent.
From \eqref{Extremal}, one might expect that the differential order of \eqref{FuncShape1} and \eqref{FuncShape2} is four.
While correct in the present case, the general situation is more complicated, as \eqref{FuncShape1}, \eqref{FuncShape2} is not a system of ordinary differential equations (ODE) but a system of differential-algebraic equations (DAE), which involves algebraic constraints.
The DAE structure is made manifest be rewriting  \eqref{FuncShape1}, \eqref{FuncShape2} as system of first order equations. 
The differential order $D$ of a DAE with $M$ first-order ODEs, $N$ constraints and differential index $I$ is \cite{Ascher1998},
\begin{align}
D=M-(I-1)N.\label{DAI}
\end{align}
The differential index $I$ is defined as the minimum number of differentiations required to write the  DAEs as ODEs.
For the EOMs  \eqref{FuncShape1}, \eqref{FuncShape2}, we find $M=7$, $N=1$ and $I=4$.
Hence, $D=4$ and a general solution of \eqref{FuncShape1}, \eqref{FuncShape2} will generically depend on four integration constants.

Following the treatment of \cite{Lue2015} in the context of QG, for the numerical analysis it is more convenient to work with a system of two second-order EOMs
\begin{align}
E^{tt}(B_2,B_1,B, A_1, A,r;M_{\mathrm{S}})={}&0,\label{FuncShapeB1}\\
E^{rr}(B_2,B_1,B, A_2,A_1,A,r;M_{\mathrm{S}}  )={}&0.\label{FuncShapeB2}
\end{align}
The $E^{tt}$ and $E^{rr}$ equations are defined as\footnote{Modulo possible singularities due to zeros in the denominators of $X$, $Y$ and $Z$, the systems \eqref{FuncShape1}, \eqref{FuncShape2} and  \eqref{FuncShapeB1}, \eqref{FuncShapeB2} are equivalent.}   
\begin{align}
E^{tt}:={}&\mathscr{E}^{tt}-X\mathscr{E}^{rr}-Y\partial_r\mathscr{E}^{rr},\\
E^{rr}:={}&\mathscr{E}^{rr}-Z\partial_{r}E^{tt}.\label{NewEOM}
\end{align}
With $a_1:=r\partial_r\ln A$, $b_1:=r\partial_r\ln B$, $b_2:=r\partial_rb_1$, we have
\begin{align}
X:={}&-\frac{B}{A}\left[1+\frac{24-2(2+a_1)(4+b_1)}{\left(4+b_1\right)^2}\right],\\
Y:={}&-2Br\left[A(4+b_1)\right]^{-1},\\
Z:={}&\left[A (b_1+4)^3 r\right]\left\{6 B \left[2 b_1 \left(-A M_{\mathrm{S}}^2
	r^2+b_1+b_2+2\right)\right.\right.\nonumber\\
	&\left.\left.+4 \left(b_2-2 A M_{\mathrm{S}}^2 r^2\right)-a_1
	(b_1+4) b_1+b_1^3\right]\right\}^{-1}.
\end{align} 
The explicit expressions for $E^{tt}$ and $E^{tt}$ are presented in \eqref{EEttExpl} and \eqref{EErrExpl}.
\vspace{-6mm}

%%%%%%%%%%%%%%%%%%%%%%%%%%%%%%%%%%%%%%%
\section{Solutions in Starobinsky's model}
\label{Sec:SolInStaro}
%%%%%%%%%%%%%%%%%%%%%%%%%%%%%%%%%%%%%%%

Even the reduced system of equations \eqref{FuncShapeB1}, \eqref{FuncShapeB2} is too complicated to be solved analytically.
An obvious strategy is to make additional assumptions, such as e.g.~a constant scalar curvature $R=R_0$, a particular relation between $B$ and $A$, or a specific ansatz for the functions $B$ and $A$. 
While such assumptions lead to strong simplifications and, in some cases, admit exact analytical solutions, they do not provide a general strategy to systematically explore the space of solutions.
In this letter, we do not make any additional simplifying assumptions, but instead classify different solutions according to their asymptotic behaviour at small and large radii and then connect these two regimes by solving the system \eqref{FuncShapeB1}, \eqref{FuncShapeB2} numerically. 
Various special radii $r$, corresponding to relevant regimes in the problem, are listed in Table \ref{Table1}.
\begin{table}[h!]
	\begin{center}
		\begin{tabular}
			{ll}
			\toprule
			$r=0$ & Spatial origin  of SS coordinates $(t,r,\vartheta,\phi)$\\
			\midrule
			$0\leq r\leq r_{\mathrm{F}}$& Frobenius regime $A(r)\sim r^s$, $B(r)\sim r^t$\\
			\midrule		
			$r_{\mathrm{S}}=M_{\mathrm{S}}^{-1}$ & Compton wavelength of the scalaron \\
			\midrule
			$r_{\mathrm{L}}\leq r\leq\infty$& Linearised regime $|W(r)|\ll1$, $|V( r)|\ll1$\\
			\midrule
			$r=\infty$& Asymptotic flatness $A(\infty)=B(\infty)\sim1$\\
			\bottomrule
		\end{tabular}
		\caption{Various regimes and scales for different SS radii $r$. The Frobenius expansions of $A(r)$ and $B(r)$ are defined in \eqref{Frob1} and \eqref{Frob1}. The functions $W(r)$ and $V(r)$ are defined below in \eqref{DefVW}.}
		\label{Table1}
	\end{center}
\end{table}
\vspace{-6mm}

%%%%%%%%%%%%%%%%%%%%%%%%%%%%%%%%%%%%%%%
\subsection{Asymptotics at the origin}
\label{Sec:AsymptOrigin}
%%%%%%%%%%%%%%%%%%%%%%%%%%%%%%%%%%%%%%%

Following the strategy of \cite{Stelle1978}, applied in the context of QG in \cite{Lu2015,Lue2015,Perkins2016-09,Holdom2017,Podolsky2018,Podolsky2020}, the solutions can be systematically classified by expanding $B$ and $A$ in a Frobenius series around $r=0$,
\begin{align}
B(r)={}&b_{t}\left(r^t+b_{t+1}r^{t+1}+\ldots\right)\label{Frob1},\\
A(r)={}&a_{\mathrm{s}}r^s+a_{s+1}r^{s+1}+\ldots\,.\label{Frob2}
\end{align} 
In \eqref{Frob1}, $b_t$ has been factored out, as it can be absorbed by rescaling the time coordinate ${t\to\tilde{t}=\sqrt{b_{t}}t}$.
Consistent combinations of $(s,t)$ follow from the \textit{indicial} equations, which are derived by inserting \eqref{Frob1}, \eqref{Frob2} into \eqref{FuncShapeB1}, \eqref{FuncShapeB2}.
We find three subcases:
\begin{align}
s<0:\quad &\text{no solution},\label{NoSol}\\
s=0:\quad  &t\text{ arbitrary},\label{szero}\\
s>0:\quad &s=(t^2+2t+4)(t+4)^{-1}.\label{sgtr}
\end{align} 
For $s<0$, no solution exists.
For $s=0$, the value of $t$ is unconstrained and there are infinitely many solutions with fixed $a_0=(1+t/2+t^2/4)$.
For $s>0$ and $s,t\in\mathbb{R}$, there are infinitely many solutions of \eqref{sgtr}.
For $s>0$ and $s,t\in\mathbb{Z}$, there are only six different integer families $(s,t)$ which satisfy \eqref{sgtr}.\footnote{In QG, only the $(0,0)$, $(-1,1)$ and $(2,2)$ families were found \cite{Lue2015}.}
The numerical analysis shows that, in contrast to the $s>0$ families, the $s=0$ families cannot be connected to asymptotically flat solutions of the full non-linear equations, except for $t=0$.
Therefore, there are a total of seven asymptotically flat, integer Frobenius families $(s,t)$ summarized in Table \ref{Table2}.
In the limit $r\to0$ for $s\neq0$, the leading $r$-dependence of the Kretschmann scalar ${K:=R_{\mu\nu\rho\sigma}R^{\mu\nu\rho\sigma}\propto r^{-2(2+s)}}$ implies that theses solutions have a singular geometry.\footnote{The $(0,0)$ family is special.
Extracting the Frobenius coefficients in \eqref{Frob1}, \eqref{Frob2} up to $\mathcal{O}(r^8)$, leads to two solutions ${\underset{r\to0}{\lim}K= (2M_{\mathrm{S}}^4/3) \left[4+\tilde{a}_{2}(1+\tilde{a}_{2})\mp2\left(4+\tilde{a}_2(2+\tilde{a}_{2})\right)^{1/2}\right]}$ with $\tilde{a}_2:=6a_2/M_{\mathrm{S}}^2$ a free parameter. 
For $\tilde{a}_{2}=0$, the minus sign (corresponding to Minkowski space) implies ${\underset{r\to0}{\lim}K=0}$. 
}
\begin{table}[h!]
		\begin{center}
		\begin{tabular}
			{llllllll}
			\toprule	
			$(s,t)$ & (0,0) & (1,-1) & (1,0) & (2,2) & (2,-2) & (7,-3)& (7,8) \\ \midrule
			$\underset{r\to0}{\lim}K$ & $\text{const.}$ & $r^{-6}$ & $r^{-6}$ &	$r^{-8}$ & $r^{-8}$ & $r^{-18}$ & $r^{-18}$\\
			\bottomrule	
		\end{tabular}
		\caption{First row: integer $(s,t)$ Frobenius families. Second row: leading $r$ scaling of the Kretschmann scalar $K$ for $r\to0$.}
		\label{Table2}
		\end{center}
\end{table}
\vspace{-6mm}

The vacuum solutions of GR are also solutions of \eqref{Staract}.
In particular, Minkowski space (MS) is included in the $(0,0)$ family, while the Schwarzschild solution (SS) is included in the $(1,-1)$ family.\vspace{2mm}

%%%%%%%%%%%%%%%%%%%%%%%%%%%%%%%%%%%%%%%
\subsection{Linearised solutions and asymptotic flatness}
\label{Sec:LinSolAsymptInfty}
%%%%%%%%%%%%%%%%%%%%%%%%%%%%%%%%%%%%%%%

The linearised EOMs in the weak field regime $r\geq r_\mathrm{L}$ are derived by inserting the decomposition
\begin{align}
B(r)=1+V(r),\qquad A(r)={}1+W(r),\label{DefVW}
\end{align}
into \eqref{FuncShapeB1},\eqref{FuncShapeB2} and by keeping only terms linear in $V$ or $W$.
The linearised EOMs have the analytic solution \cite{Pechlaner1966,Stelle1978},
\begin{align}
V(r)={}&C+\frac{C_{2}}{r}+\sum_{\sigma=-,+}\frac{C_{0}^{\sigma}e^{\sigma M_{\mathrm{S}}r}}{r}, \label{linsol1}\\
W(r)={}&-\frac{C_{2}}{r}+\sum_{\sigma=-,+}\frac{C_{0}^{\sigma}e^{\sigma M_{\mathrm{S}}r}}{r}\left(1-\sigma M_{\mathrm{S}}r\right).\label{linsol2}
\end{align}
The four integration constants $C$, $C_{2}$, and $C_{0}^{\pm}$ agree with the result $D=4$ found 
by the DAE analysis \eqref{DAI}.\footnote{The same DAE analyses in QG and EWG lead to ${D=6}$ and ${D=4}$, matching the number of integration constants of the linearised equations, i.e.~twice the number of propagating particles.}
Asymptotic flatness of the geometry requires
\begin{align}
\lim_{r\to\infty}B(r)=1,\qquad \lim_{r\to\infty}A(r)=1.\label{AScon}
\end{align}
The limits \eqref{AScon}, in turn, require the constant $C$ and the constant ${C_{0}^{+}}$ of the rising scalaron Yukawa potential to vanish, reducing the number of free parameters to two, 
\begin{align}
V^{\infty}(r):={}&\frac{C_{2}}{r}+\frac{C_{0}^{-}e^{-M_{\mathrm{S}}r}}{r},\label{linsol1af}\\
W^{\infty}(r):={}&-\frac{C_{2}}{r}+\frac{C_{0}^{-}e^{-M_{\mathrm{S}}r}}{r}\left(1+M_{\mathrm{S}}r\right).
\label{linsol2af}
\end{align}
Matching \eqref{linsol1af} with the Newtonian potential at $r\to\infty$, 
\begin{align}
\lim_{r\to\infty}V^{\infty}(r)=-2\frac{G_{\mathrm{N}}M_{\mathrm{T}}^{\infty}}{r},\label{Newton}
\end{align}
relates $C_2$ with the total mass $M_{\mathrm{T}}^{\infty}$ at infinity.
In terms of Planck units, this implies $C_2=-16\pi M_{\mathrm{T}}^{\infty}$.
In contrast, $C_0^{-}$ cannot be directly related to an observable at infinity and a priori remains an undetermined parameter.\vspace{2mm}

%%%%%%%%%%%%%%%%%%%%%%%%%%%%%%%%%%%%%%%
\subsection{Numerical algorithm}
\label{Sec:NumAlgo}
%%%%%%%%%%%%%%%%%%%%%%%%%%%%%%%%%%%%%%%

We connect the asymptotics of the solutions $A$ and $B$ in the Frobenius regime with those in the linearised regime, by numerically solving the full non-linear equations \eqref{NewEOM}.
This is a boundary problem which is difficult to solve -- even numerically.
Therefore, we use a double-shooting algorithm.
We first formulate the problem as initial value problem (IVP) with ``asymptotic initial conditions'' generated by matching the full numerical solution to the linearised solution at large radii $r_{\mathrm{I}}\geq r_{\mathrm{L}} $, 
\begin{align}
B(r_{\mathrm{I}})={}&1+V^{\infty}(r_{\mathrm{I}}),\qquad B_1(r)|_{r_{\mathrm{I}}}={}V^{\infty}_1(r)|_{r_{\mathrm{I}}},\\  A(r_{\mathrm{I}})={}&1+W^{\infty}(r_{\mathrm{I}}),\qquad A_1(r)|_{r_{\mathrm{I}}}=W^{\infty}_1(r)|_{r_{\mathrm{I}}}.
\end{align}
For fixed $M_{\mathrm{S}}$, the IVP only depends on the two parameters $C_0^{-}$ and $C_2$.
Fixing $C_0^{-}$ and $C_2$, we integrate the system inwards from the initial radius $r_{\mathrm{I}}$ to the final radius $r_{\mathrm{E}}$ at which the solutions $A$ and $B$ are dominated by the leading term of the Frobenius series.
Matching the numerical solutions $A$ and $B$ at $0<r_{\mathrm{E}}\leq r_{\mathrm{F}}$ to \eqref{Frob1} and \eqref{Frob2}, the numerical values $(s_{\mathrm{N}},t_{\mathrm{N}})$ are extracted from the slope of a linear log-log fit with constants $c_B$ and $c_A$,
\begin{align}
\ln B(r_{\mathrm{E}}) ={}&c_B+ t_{\mathrm{N}} \ln r_{\mathrm{E}}+\ldots,\label{LogFit1} \\
\ln A(r_{\mathrm{E}})  ={}&c_A+ s_{\mathrm{N}} \ln r_{\mathrm{E}} +\ldots,\label{LogFit2}
\end{align}
In this way, for each pair of initial values $(C_2,C_0^{-})$, a pair of numerical values $(s_{\mathrm{N}},t_{\mathrm{N}})$ is found.
Repeating this procedure by scanning over the $(C_2,C_0^{-})$ parameter space, we only store those values of $(C_2,C_0^{-})$ for which the numerical values $(s_{\mathrm{N}},t_{\mathrm{N}})$ coincide with the $(s,t)$ values of one of the integer Frobenius families up to a tolerance $\max\{\Delta s,\Delta t\}< \delta_{\mathrm{S}}=10^{-3}$, with the fractional differences  $\Delta s:=(s-s_\mathrm{N})/s$ and $\Delta t:=(t-t_\mathrm{N})/t$, between the corresponding numerical values $(s_{\mathrm{N}},t_{\mathrm{N}})$ and the integer Frobenius values $(s,t)$.

By construction of the shooting algorithm, the numerical solutions $A$ and $B$ are only strictly valid up to $r_{\mathrm{I}}$.
Despite the fact that $M_{\mathrm{S}}$ parametrically enters the EOMs \eqref{FuncShapeB1},\eqref{FuncShapeB2}, from the point of view of the initial conditions \eqref{linsol1af},\eqref{linsol2af}, the value of $r_{\mathrm{I}}$ is tied to the value of $M_{\mathrm{S}}$, which we choose according to the CMB constraints \eqref{CMBBoundMS}, i.e.~in Planck units $M_{\mathrm{S}}=10^{-5}$.
For a given $r_{\mathrm{I}}$, different ``confidence regions'' in $(C_2,C_{0}^{-})$ space correspond to different tolerance values ${\max\{|V^{\infty}(r_{\mathrm{I}})|,|W^{\infty}(r_{\mathrm{I}})|\}<\delta_{\mathrm{L}}}$, which quantify how well the linearised approximation is satisfied at $r_{\mathrm{I}}$.
While a large $r_{\mathrm{I}}$ would be desirable as it allows for larger values of $(C_2,C_{0}^{-})$ within the same confidence interval, it is computationally expensive.
For the numerical analysis, we choose $r_{\mathrm{I}}=10$ (in Planck units).
Likewise, choosing $r_{\mathrm{E}}$ too small, the singularity of the EOMs at $r=0$ might cause the solution to diverge already for $r>r_{\mathrm{E}}$, while choosing $r_{\mathrm{E}}$ too large might result in a poor fit \eqref{LogFit1}, \eqref{LogFit2}.
We choose the default value $r_{\mathrm{E}}=10^{-4}$, but if the quality of the fit falls below a tolerance $\delta_{F}$, the code dynamically adjusts by automatically lowering the value of $r_{\mathrm{E}}$. 

We performed various consistency checks.
First, we repeated the Frobenius analysis also in Kundt coordinates (as done in \cite{Podolsky2018,Podolsky2020} for QG) and found the same integer Frobenius families as in SS coordinates.
Second, as a check of our numerical implementation, we applied our algorithm to the model of QG and found good agreement with the numerical results obtained in \cite{Perkins2016-09}.
Third, we checked that the numerical solutions of the reduced system \eqref{FuncShapeB1},\eqref{FuncShapeB2} also satisfy the original system \eqref{FuncShape1}, \eqref{FuncShape2}, by re-substituting them into the original system.
Fourth, we checked that for all (including non-integer) numerically found values $(s_{\mathrm{N}},t_{\mathrm{N}})$, the Frobenius constraint \eqref{sgtr} is satisfied with a tolerance not exceeding ${|s-(t^2+2t+4)(t+4)^{-1}|<\delta_\mathrm{C}=10^{-3}}$.\footnote{The numerically extracted $(s,t)$ values, all satisfying \eqref{sgtr}, vary smoothly under variation of $(C_2,C_{0}^{-})$, implying that non-integer families are qualitatively not different from integer ones.}\vspace{2mm}

%%%%%%%%%%%%%%%%%%%%%%%%%%%%%%%%%%%%%%%
\section{Results}
\label{Sec:ResCon}
%%%%%%%%%%%%%%%%%%%%%%%%%%%%%%%%%%%%%%%

\begin{figure}[h!]
	\includegraphics[width=8cm,height=4cm]{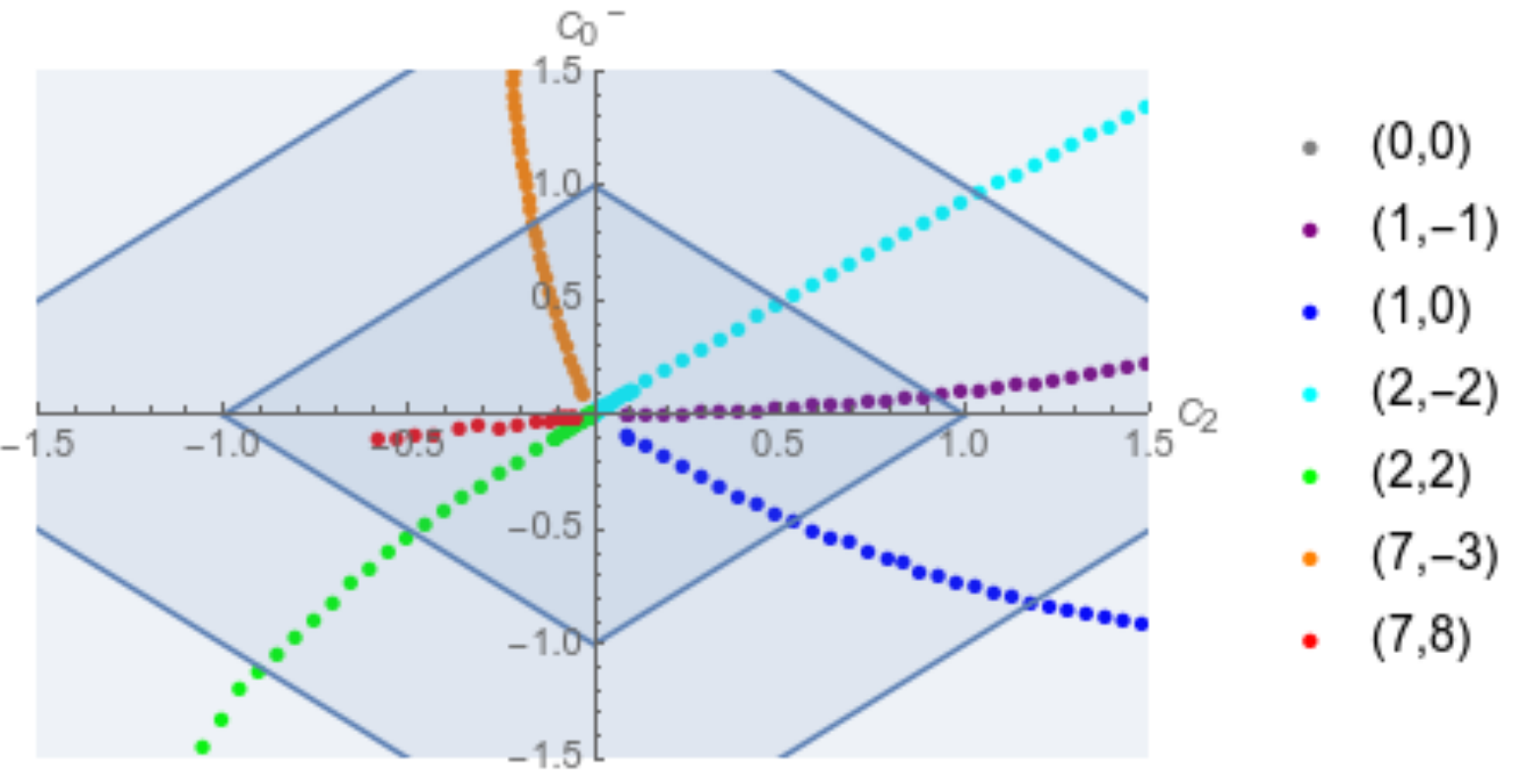}
	\caption{All integer Frobenius families trace out a line in the $(C_2,C_0^{-})$ parameter space, except for the $(0,0)$ family, which leads to a (not visible) single point at the origin.
	The coloured areas correspond to (inwards to outwards) $\delta_{\mathrm{L}}<\{0.1,0.2,0.3\}$ quantifying the quality of the linearised approximation at $r_{\mathrm{I}}=10$. The boundary $\delta_{\mathrm{L}}=0.3$ is not visible.
	Numerical difficulties prevent generating $(7,8)$ points beyond $C_{2}<-0.6$.}
	\label{Fig1}
\end{figure}

We find that all integer Frobenius families can be connected to asymptotically flat solutions and lead to the ``phase-diagram'' Fig.~\ref{Fig1}.

Except for the $(0,0)$ family, which appears as a single point at the origin in Fig.~\ref{Fig1}, all integer Frobenius families trace out lines in parameters space, which emanate in the vicinity of the origin and expand outwards to larger $(C_2,C_{0}^{-})$ values without intersecting. While Minkowski space is contained in the $(0,0)$ family, the Schwarzschild solution, which is formally part of the $(1,-1)$ family, is not contained in the numerically found solutions because it has a horizon.\footnote{Numerically, we could neither find solution with $s<0$, confirming the expected result from \eqref{NoSol}, nor could we find any of the $(0,t)$ except for the $(0,0)$ family, suggesting that the $(0,t)$, $t\neq0$ families cannot be connected to asymptotically flat solutions.
The $(0,0)$ family appears as a single point in Fig.~\ref{Fig1}, as to all orders \eqref{Frob1}, \eqref{Frob2} depends on a single Frobenius parameter.}

According to Table \ref{Table2}, all integer Frobenius families (except the $(0,0)$ family) have a singular geometry.
By construction, none of the found solution has a horizon, such that the corresponding solutions all feature a \textit{naked singularity} at the origin.\footnote{In the shooting algorithm solutions with horizons are discarded as they would diverge at some $r_{\mathrm{H}}>r_{\mathrm{E}}$ such that the numerical integration would fail before a successful fit of $(s_{\mathrm{N}},t_{\mathrm{N}})$ could be obtained.
In any case, the uniqueness theorem for Starobinsky's model guarantees that the SS is the only spherically symmetric, asymptotically flat vacuum solution with a horizon \cite{Whitt1984}.
Nevertheless, by adapting the shooting algorithm, making it capable of detecting a horizon, we numerically ``confirm'' this, as, similar to the findings in \cite{Perkins2016-09}, the only positive mass solution with horizon we found is the SS solution lying on the negative $C_{2}$ axis in Fig.~\ref{Fig1}.
}

Since, according to \eqref{Newton}, $C_2$ is connected to the total mass at infinity, positivity of $M_{\mathrm{T}}^{\infty}$ requires $C_2$ to be negative.
The physical significance of the negative mass families $(1,-1)$, $(1,0)$ and $(2,-2)$ remains unclear, but we focus on the positive mass families $(2,2)$, $(7,-3)$ and $(7,8)$ which lie in the negative $C_2$ half-plane of Fig.~\ref{Fig1}. 
\begin{figure}[]
	\includegraphics[width=8.3cm,height=12cm]{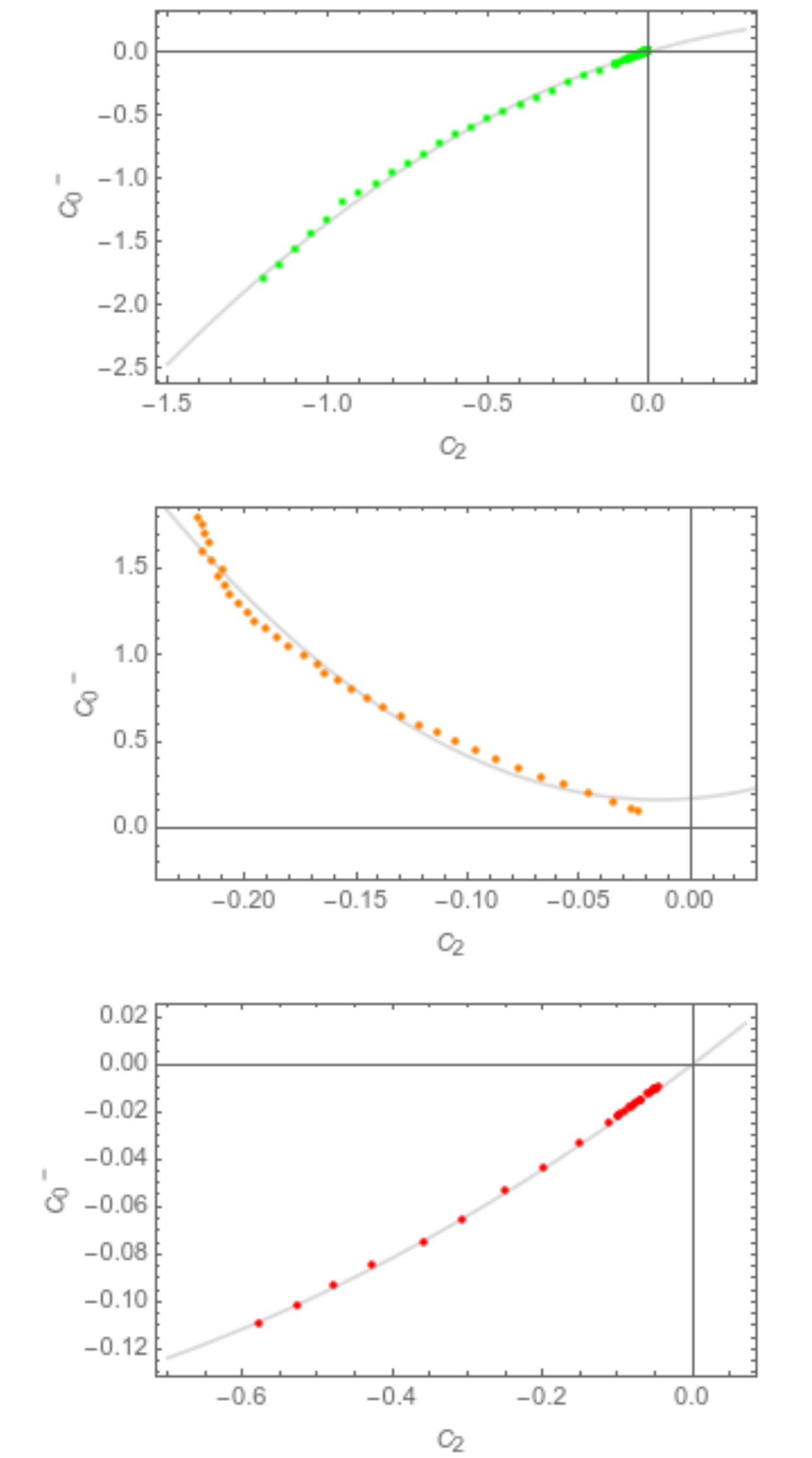}
	\caption{Quadratic fits of the positive mass families (the colour coding is the same as in Fig.~\ref{Fig1}). Upper diagram: quadratic fit $C_0^{-}=0.76 C_2 - 0.59 C_2^2$ of the $(2,2)$ line in Fig.~\ref{Fig1}, valid for $-1.2<C_2<0$ for which $\delta_{L}<0.2$.
	Central diagram: quadratic fit $C_0^{-}=0.17 + 0.99 C_2 + 34.45 C_2^2$ of the $(7,-3)$ line in Fig.~\ref{Fig1}, valid for $-0.22<C_2<0$ for which $\delta_{L}<0.2$.
	Lower diagram: quadratic fit $C_0^{-}=0.24 C_2 + 0.09 C_2^2$ of the $(7,8)$ line in Fig.~\ref{Fig1}, valid for $-0.6<C_2<0$ for which $\delta_{L}<0.2$.}
	\label{Fig2}
\end{figure}

In contrast to $C_2$, the parameter $C_0^{-}$ cannot directly be related to an observable at infinity.
However, the phase diagram Fig.~\ref{Fig1} shows that for each integer Frobenius family, $C_0^{-}$ can uniquely be expressed as a function of $C_2$ and therefore as a function of $M_{\mathrm{T}}^{\infty}$. 
Within the reliable $C_{2}$ intervals of each family, the function $C_{0}^{-}(C_2)$ is well described by a quadratic fit shown in Fig.~\ref{Fig2} for the positive mass families.
This implies that all solutions in the integer Frobenius families are completely characterized by two $(s,t)$ values and by one single physical parameter $M_{\mathrm{T}}^{\infty}$.

In the numerical analysis we assumed $M_{\mathrm{S}}=10^{-5}$ and $r_{\mathrm{I}}=10$, such that, even for radii $r\geq r_{\mathrm{L}}$ well within the linearised regime, we can expand the exponential in \eqref{linsol1af},
\begin{align}
V^{\infty}(r)=\frac{C_2+C_0^{-}}{r}+\mathrm{O}\left(M_{\mathrm{S}}r\right),\qquad r\leq r_{\mathrm{S}},
\end{align} 
which implies the constraint $(C_{2}+C_{0}^{-})/r_{\mathrm{I}}<\delta_{\mathrm{L}}$.
Using the quadratic fits obtained in Fig.~\ref{Fig2} with the identification $C_{2}=-16 M_{\mathrm{T}}^{\infty}$, we encounter the interesting situation that even in the linearised regime, the ``effective quasi-local mass'' $C_{2}+C_{0}^{-}$ might vanish or become negative, rendering the weak field potential $V^{\infty}(r)$ repulsive.
For the $(2,2)$ family, we obtain in this way
\begin{align}
V^{\infty}(r)\approx-\frac{16\pi M_{\mathrm{T}}^{\infty}}{r}\left(1.76-0.56 M_{\mathrm{T}}^{\infty}\right),\label{Screening}
\end{align} 
which has a zero at $M_{\mathrm{T}}^{\infty}\approx\pi$ and even becomes negative for larger values of $M_{\mathrm{T}}^{\infty}$.
This scalaron-induced ``mass screening mechanism'' for $r<r_{\mathrm{S}}$ might have interesting implications for the  solutions, as e.g.~avoiding a repulsive potential would imply a constraint on the total mass at infinity $M_{\mathrm{T}}^{\infty}<\pi$, which, however, in the present case for $M_{\mathrm{T}}^{\infty}=\pi$ lies outside the region of validity of the quadratic fit of the $(2,2)$ family as ${-16\pi^2=C_2\ll-1.2}$, cf.~Fig.~\ref{Fig2}.
Similar considerations for the $(7,-3)$ and $(7,8)$ families would imply an anti-screening. 

Summarizing, we found new static, spherically symmetric and asymptotically flat vacuum solutions of Starobinsky's quadratic $f(R)$ model which feature naked singularities.
The phase diagram Fig.~\ref{Fig1} might be interpreted as  ``Frobenius no-hair theorem'' as all solutions in the seven integer Frobenius families labelled by the two Frobenius indices $(s,t)$ are characterized by one single physical parameter -- the total mass at infinity $M_{\mathrm{T}}^{\infty}$.
Besides the condition of positive total mass at infinity, we found a mass (anti)screening mechanism, which depends on the Frobenius family and the scalaron mass $M_{\mathrm{S}}$, and might lead to additional constraints on the Frobenius solutions.  
The implications of the naked singularity solutions in the context of the black hole information paradox requires a more detailed study involving the analysis of the shape of different solutions within a given Frobenius family. We hope to address this question in a forthcoming work.
Finally, it would be interesting to apply the presented methods to other relevant $f(R)$ models and modifications of GR.

%%%%%%%%%%%%%%%%%%%%%%%%%%%%%%%%%%%%%%%%
\appendix
\onecolumngrid
%%%%%%%%%%%%%%%%%%%%%%%%%%%%%%%%%%%%%%%
\section{Explicit form of EOMs}
\label{App:ExplEOMs}
The explicit expressions for the coefficients $\mathscr{E}^{tt}$ and $\mathscr{E}^{rr}$ of \eqref{Extremal}, which enter the EOMs \eqref{FuncShape1} and \eqref{FuncShape2}, read
\begin{align}
\mathscr{E}^{tt}={}&
\frac{M_{\mathrm{P}}^2}{96 M_{\mathrm{S}}^2 r^4
	A^5 B^3} \left\{A^3 \left[16 B^4 \left(3 M_{\mathrm{S}}^2 r^3 A_1+5\right)-4 r^2 B^2 \left(4
	r B_1 \left\{11 B_2+3 r B_3\right\}+4 B_1^2+9 r^2 B_2^2\right)\right.\right.\nonumber\\
	&\left.+4 r^3 B
	B_1^2 \left(26 B_1+29 r B_2\right)-49 r^4 B_1^4+16 r^3 \left(4 B_3+r
	B_4\right) B^3\right]-2 r A^2 B \left[4 r B^2 \left(r \left\{\left[6 A_2+r
	A_3\right] B_1\right.\right.\right.\nonumber\\
	&\left.\left.+4 r A_2 B_2\right\}-2 A_1 \left\{B_1-r \left(8 B_2+3 r
	B_3\right)\right\}\right)-6 r^2 B B_1 \left(2 B_1 \left\{5 A_1+r A_2\right\}+9 r
	A_1 B_2\right)+29 r^3 A_1 B_1^3\nonumber\\
	&\left.+16 B^3 \left(r \left\{A_2+r A_3\right\}-2
	A_1\right)\right]+r^2 A B^2 A_1 \left[4 r B \left(B_1 \left\{28 A_1+13 r
	A_2\right\}+19 r A_1 B_2\right)\right.\nonumber\\
	&\left.-57 r^2 A_1 B_1^2+16 B^2 \left(5 A_1+13 r
	A_2\right)\right]-56 r^3 B^3 A_1^3 \left[r B_1+4 B\right]+16 A^5 B^4 \left[3
	M_{\mathrm{S}}^2 r^2+1\right]\nonumber\\
	&\left.-48 A^4 B^4 \left[M_{\mathrm{S}}^2 r^2+2\right]\right\},\label{EttExpl}\\
\mathscr{E}^{rr}={}&
\frac{M_{\mathrm{P}}^2}{96 M_{\mathrm{S}}^2 r^4 A^3 B^4} \left\{2 r^2 A B \left[r B_1+4 B\right] \left[2 B \left(B_1 \left(4
	A_1+r A_2\right)+2 r A_1 B_2\right)-3 r A_1 B_1^2+8 B^2 A_2\right]\right.\nonumber\\
	&-7 r^2
	B^2 A_1^2 \left[r B_1+4 B\right]^2+48 A^3 B^3 \left[M_{\mathrm{S}}^2 r^3 B_1+B
	\left(M_{\mathrm{S}}^2 r^2-2\right)\right]+A^2 \left[-32 r B^3 \left(r \left\{2 B_2+r
	B_3\right\}-2 B_1\right)\right.\nonumber\\
	&\left.+4 r^2 B^2 \left(-2 r B_1 \left\{r B_3-6
	B_2\right\}+16 B_1^2+r^2 B_2^2\right)+4 r^3 B B_1^2 \left(3 r B_2-4
	B_1\right)-7 r^4 B_1^4+112 B^4\right]\nonumber\\
	&\left.-16 A^4 B^4 \left[3 M_{\mathrm{S}}^2
	r^2+1\right]\right\}\label{ErrExpl}.
\end{align}
The explicit expressions for the coefficients $E^{tt}$ and $E^{rr}$, which enter the EOMs \eqref{FuncShapeB1} and \eqref{FuncShapeB2}, read
\begin{align}
E^{tt}={}&
\frac{M_{\mathrm{P}}^2}{8 M_{\mathrm{S}}^2 r^4 A^4 B^2 \left\{r B_1+4
	B\right\}^2}\nonumber\\
	&\times \left\{A^2 \left[-4 r^2 B^3 \left(B_1^2 \left\{M_{\mathrm{S}}^2 r^3 A_1+4\right\}+8 r
	B_1 B_2+2 r^2 B_2^2\right)+8 r B^4 B_1 \left(r \left\{2-M_{\mathrm{S}}^2 r^2\right\}
	A_1+2\right)+32 B^5 \left(r \left\{M_{\mathrm{S}}^2 r^2\right.\right.\right.\right.\nonumber\\
	&\left.\left.\left.+2\right\} A_1+1\right)-4 B^2 \left(r^5
	B_1 B_2^2-4 r^3 B_1^3\right)+2 r^4 B B_1^3 \left(B_1+2 r B_2\right)-r^5
	B_1^5\right]-2 r A B A_1 \left[r B_1+4 B\right] \left[-2 r^2 B B_1
	\left(B_1\right.\right.\nonumber\\
	&\left.\left.+r B_2\right)+r^3 B_1^3+8 B^3\right]+r^2 B^2 A_1^2 \left[2 B-r
	B_1\right] \left[r B_1+4 B\right]^2-4 A^3 B^2 \left[2 r^2 B B_1 \left(\left\{3
	M_{\mathrm{S}}^2 r^2+2\right\} B_1-M_{\mathrm{S}}^2 r^3 B_2\right)\right.\nonumber\\
	&\left.-8 B^2 \left(r \left\{M_{\mathrm{S}}^2
	r^2-1\right\} B_1+M_{\mathrm{S}}^2 r^4 B_2\right)+M_{\mathrm{S}}^2 r^5 B_1^3+8 B^3 \left(M_{\mathrm{S}}^2
	r^2+2\right)\right]+16 A^4 B^4 \left[r \left(2 M_{\mathrm{S}}^2 r^2+1\right) B_1\right.\nonumber\\
	&\left.\left.+2 B
	\left(M_{\mathrm{S}}^2 r^2+1\right)\right]\right\},\label{EEttExpl}\\
E^{rr}={}&
-\frac{M_{\mathrm{P}}^2}{96
	M_{\mathrm{S}}^2 A^3 B^4 r^5 \left\{2 M_{\mathrm{S}}^2 A^2 r \left[4 B+r B_1\right] B^2+r
	A_1 B_1 \left[4 B+r B_1\right] B+A \left[r^2 B_1^3-8 B^2 B_1-2 B
	r \left(2 B+r B_1\right) B_2\right]\right\}}\nonumber\\
	&\times \left\{32 M_{\mathrm{S}}^2 A^6 r^2 \left[3 M_{\mathrm{S}}^2 r^2+1\right] \left[4 B+r
	B_1\right] B^6+16 A^5 \left[8 \left(-3 M_{\mathrm{S}}^4 r^4+10 M_{\mathrm{S}}^2 r^2+8\right) B^3+2 r
	\left(\left\{-15 M_{\mathrm{S}}^4 r^4+2 M_{\mathrm{S}}^2 r^2\right.\right.\right.\right.\nonumber\\
	&
	\left.\left.\left.\left.+12\right\} B_1-2 r \left\{5 M_{\mathrm{S}}^2 r^2+1\right\}
	B_2\right) B^2-2 r^2 B_1 \left(M_{\mathrm{S}}^2 B_2 r^3+\left\{3 M_{\mathrm{S}}^4 r^4+1\right\}
	B_1\right) B-r^3 \left(5 M_{\mathrm{S}}^2 r^2+3\right) B_1^3\right] B^4\right.\nonumber\\
	&
	\left.+r^3 A_1^3 \left[16
	B-r B_1\right] \left[4 B+r B_1\right]^3 B^3-A r^2 A_1 \left[4 B+r
	B_1\right]^2 \left[32 \left(2 A_1+r A_2\right) B^3+4 r \left(2 B_1 \left\{9 A_1+r
	A_2\right\}\right.\right.\right.\nonumber\\
	&\left.\left.+5 r A_1 B_2\right) B^2-2 r^2 A_1 B_1 \left(B_1+2 r B_2\right)
	B-r^3 A_1 B_1^3\right] B^2-2 A^4 \left[-7 M_{\mathrm{S}}^2 B_1^5 r^7+8 B B_1^3
	\left(2 M_{\mathrm{S}}^2 r^3 B_2\right.\right.\nonumber\\
	&\left.-\left\{M_{\mathrm{S}}^2 r^2+2\right\} B_1\right) r^4+4 B^2 B_1
	\left(-M_{\mathrm{S}}^2 B_2^2 r^4+6 M_{\mathrm{S}}^2 B_1 B_2 r^3+\left\{54 M_{\mathrm{S}}^2 r^2+4\right\}
	B_1^2\right) r^3+8 B^3 \left(-2 M_{\mathrm{S}}^2 B_2^2 r^4\right.\nonumber\\
	&\left.-34 M_{\mathrm{S}}^2 B_1 B_2
	r^3-\left\{8 M_{\mathrm{S}}^2 r^2+\left[3 M_{\mathrm{S}}^2 r^3+r\right] A_1-12\right\} B_1^2\right) r^2+16
	B^4 \left(\left\{3 M_{\mathrm{S}}^2 r^2-2 \left(3 M_{\mathrm{S}}^2 r^3+r\right) A_1+56\right\} B_1\right.\nonumber\\
	&\left.\left.+2 r
	\left\{6-5 M_{\mathrm{S}}^2 r^2\right\} B_2\right) r+64 B^5 \left(11 M_{\mathrm{S}}^2 r^2+16\right)\right]
	B^2+A^2 r \left[4 B+r B_1\right] \left[32 \left(A_1 \left\{r \left(11 M_{\mathrm{S}}^2
	r^2+8\right) A_1-8\right\}\right.\right.\nonumber\\
	&\left.+4 r A_2\right) B^5+16 r \left(B_1 \left\{A_1 \left[r \left(5
	M_{\mathrm{S}}^2 r^2+4\right) A_1-3\right]+10 r A_2\right\}+2 r \left\{2 r A_2-A_1\right\}
	B_2\right) B^4-2 \left(16 A_1 B_2^2 r^4\right.\nonumber\\
	&\left.-8 B_1 A_2 B_2 r^4+A_1
	\left.\{M_{\mathrm{S}}^2 r^3 A_1-48\right\} B_1^2 r^2\right) B^3-4 r^4 B_1 \left(2 A_2
	B_1^2-10 A_1 B_2 B_1+r A_1 B_2^2\right) B^2-4 r^4 A_1 B_1^3 \left(7
	B_1\right.\nonumber\\
	&\left.\left.+2 r B_2\right) B+5 r^5 A_1 B_1^5\right] B+A^3 \left[-256 \left(\left\{3
	M_{\mathrm{S}}^2 r^2+2\right\} A_2 r^2-4 A_1 r-4\right) B^7+64 r \left(B_1 \left\{r
	\left[\left(6-9 M_{\mathrm{S}}^2 r^2\right) A_1\right.\right.\right.\right.\nonumber\\
	&\left.\left.\left.-2 r \left(3 M_{\mathrm{S}}^2 r^2+2\right)
	A_2\right]+22\right\}+r \left\{r \left(3 M_{\mathrm{S}}^2 r^2+2\right) A_1+7\right\} B_2\right)
	B^6+16 r^2 B_1 \left(r^2 \left\{3 M_{\mathrm{S}}^2 r^2+2\right\} A_1 B_2\right.\nonumber\\
	&\left.-B_1 \left\{r
	\left[\left(39 M_{\mathrm{S}}^2 r^2+22\right) A_1+r \left(3 M_{\mathrm{S}}^2 r^2+2\right)
	A_2\right]-14\right\}\right) B^5-8 r^3 \left(\left\{3 r \left[3 M_{\mathrm{S}}^2 r^2+4\right]
	A_1-26\right\} B_1^3+8 r B_2 B_1^2\right.\nonumber\\
	&\left.+4 r^2 B_2^2 B_1-2 r^3 B_2^3\right) B^4+4
	r^4 B_1^2 \left(3 \left\{M_{\mathrm{S}}^2 A_1 r^3+16\right\} B_1^2+40 r B_2 B_1-2 r^2
	B_2^2\right) B^3+4 r^5 B_1^3 \left(-14 B_1^2+7 r B_2 B_1\right.\nonumber\\
	&\left.\left.\left.+3 r^2 B_2^2\right)
	B^2-2 r^6 B_1^5 \left(7 B_1+6 r B_2\right) B+3 r^7 B_1^7\right]\right\}\label{EErrExpl}.
\end{align}
\twocolumngrid
%%%%%%%%%%%%%%%%%%%%%%%%%%%%%%%%%%%%%%%%
\bibliography{SSSFRPRD}{}
%%%%%%%%%%%%%%%%%%%%%%%%%%%%%%%%%%%%%%%
\end{document}